\documentstyle[sprocl,psfig]{article}

\def\al{\alpha}

\def\ga{\gamma}
\def\de{\delta}
\def\ep{\epsilon}

\def\et{\eta}
\def\th{\theta}

\def\si{\sigma}

\def\ch{\chi}
\def\ps{\psi}

\def\Ga{\Gamma}
\def\De{\Delta}

\def\Om{\Omega}
\def\mn{{\mu\nu}}
\def\cl{{\cal L}}
\def\cC{{\cal C}}

\def\fr#1#2{{{#1} \over {#2}}}
\def\prt{\partial}

\def\pt#1{\phantom{#1}}

 \def\half{{\textstyle{1\over 2}}}

\def\frac#1#2{{\textstyle{{#1}\over {#2}}}} 
\def\lsim{\mathrel{\rlap{\lower4pt\hbox{\hskip1pt$\sim$}} 
\raise1pt\hbox{$<$}}}
\def\gsim{\mathrel{\rlap{\lower4pt\hbox{\hskip1pt$\sim$}} 
\raise1pt\hbox{$>$}}}
\def\sqr#1#2{{\vcenter{\vbox{\hrule height.#2pt 
\hbox{\vrule width.#2pt height#1pt \kern#1pt \vrule width.#2pt}
\hrule height.#2pt}}}}

\def\lrprtnu{\stackrel{\leftrightarrow}{\partial^\nu}} 
\def\lrcov{\stackrel{\leftrightarrow}{D}}

\def\lrDnuupper{\stackrel{\leftrightarrow}{D^\nu}}

\def\slash#1{\not\hbox{\hskip -2pt}{#1}}

\newcommand{\beq}{\begin{equation}}
\newcommand{\eeq}{\end{equation}}
\newcommand{\bea}{\begin{eqnarray}}
\newcommand{\eea}{\end{eqnarray}}

\begin{document}

\title{SCATTERING CROSS SECTIONS AND LORENTZ VIOLATION}

\author{DON COLLADAY}

\address{New College of Florida, 5700 Tamiami Trail, Sarasota,\\ FL
34243, USA\\E-mail: colladay@sar.usf.edu} 


\maketitle\abstracts{ To date, a significant effort has been made to
adddress the Lorentz violating standard model in low-energy systems, but
little is known about the ramifications for high-energy cross sections.  In
this talk, I discuss the modified Feynman rules that result when Lorentz
violation is present and give the results of an explicit calculation for
the process $e^+ + e^- \rightarrow 2 \gamma$.}

\section{Introduction}

Considerable attention has been paid recently to the evaluation of the
prospect of violating Lorentz invariance in a more complete theory
underlying the standard model.  The initial motivation for consideration
of  Lorentz violation came from string theory,~\cite{kps} and more
recently this breaking has been discussed in the context of noncommutative
geometry.~\cite{lane} The mechanism of spontaneous symmetry breaking has
been used to generate Lorentz-violating terms in a general, theory
independent way.  As a result, low-energy effects can be parametrized and 
understood regardless of the general structure of the underlying theory. 
A generic extension of the standard model including all observer Lorentz
scalars that are gauge invariant and power counting renormalizable has
been constructed.~\cite{ck}

A broad range of low-energy experiments have recently been performed that 
test for miniscule violations of Lorentz symmetry.~\cite{expproc} 
For example, clock comparisons, Penning trap tests and
spin-polarized torsion pendulum experiments all place
bounds on a variety of 
Lorentz-violating parameters at the order of one part in $10^{20} $ to $
10^{30}$.  So far, relatively little is known about the calculation of
cross sections or decay rates in the context of the Lorentz-violating
extension of the standard model. In this talk I will discuss the general
procedure for calculating  cross sections in the presence of Lorentz
violation.  As an explicit example I will discuss the process $e^- e^+
\rightarrow 2
\ga$ at ultrarelativistic accelerator energies.
A more detailed analysis is in the literature. \cite{collkos}

\section{Lorentz-Violating QED Extension}

To simplify the analysis, we focus here on a QED subset
of the full standard model extension. The QED extension is obtained by
restricting the full standard model extension to the electron and photon
sectors. Imposing gauge invariance and restricting to power-counting
renormalizable terms yields a lagrangian of 
\begin{equation}
\cl  =
\half i \overline{\ps}  \Ga^\nu \lrcov_\nu \ps 
- \overline{\ps}  M  \ps
+ \cl_{\rm photon}
\quad ,
\label{baselag}
\end{equation}
where $\Ga$ and $M$ are given by
\beq
{\Ga}^{\nu}  ={\ga}^{\nu}+  c^{\mu \nu} 
{\ga}_{\mu}+  d^{\mu \nu} {\ga}_{5} {\ga}_{\mu}
\quad ,
\eeq
\beq
 M  = m+  a_{\mu}  {\ga}^{\mu}+  b_{\mu}
 {\ga}_{5} {\ga}^{\mu}+\frac{1}{2} H^{\mu \nu}
 {\sigma}_{\mu \nu}
\quad .
\eeq 
The parameters $a$, $b$, $c$, $d$, and $H$ are fixed background
expectation values of tensor fields.
In this work we neglect possible Lorentz-Violating contributions to the
photon sector since these terms are stringently bounded using cosmological
birefringence tests.~\cite{km}

An immediate difficulty arises upon using the above lagrangian since
there are extra time derivative couplings arising from the $\Gamma^0$
term.
This means that the field $\psi$ does not satisfy a conventional
Schr\"odinger evolution with a hermitian hamiltonian.
To circumvent this problem, a spinor field redefinition is used to
eliminate the extra time derivatives.~\cite{bkr}
The redefinition takes the form
\beq
\psi = A \chi \quad ,
\label{fredef}
\eeq
where the matrix $A$ satisfies
\beq
\overline A \Gamma^0 A = \ga^0 \quad ;
\quad
(\overline A = \ga^0 A^\dagger \ga^0)) \quad ,
\eeq
therefore conjugating away the time derivatives.  
Such a transformation is always possible provided the observer is in a
concordant frame in which the violation parameters are small.~\cite{kle}
The redefined field obeys the standard Schr\"odinger time evolution
equation,
$i\prt_0 \ch = H \ch$,
with a modified hamiltonian given by
\beq
H = \ga^{0} \overline A ( - i \Ga_{j} D^{j} +  M) A \quad .
\eeq

\section{Modified Feynman Rules}

The general structure of the calculation of cross sections in the presence
of Lorentz violation parallels the conventional approach.
Development of the perturbative solution to a general scattering 
problem leads to a modified set of Feynman rules.
The general method of the application of the rules is similar to 
the conventional case with several important modifications.

First, translational invariance of the theory implies that $p^\mu$ is
conserved at all vertices of the diagrams as in the usual case.  
However, care must be taken to include the modified dispersion 
relation satisfied by $p^0(\vec p)$.  
This modification has an effect on the kinematics of the scattering
process and can alter the conventional particle trajectories.

Second, the spinor solutions to the modified Dirac equation must be 
included on each external leg of the diagrams.  
This fact can be deduced through application of the standard LSZ reduction
procedure for the fermions.

Third, the fermion propagator used on all internal lines of the 
diagrams takes the form

\beq
S_{F}(p) = 
\fr 1 
{\ga^{0} E - \overline A \vec \Ga A \cdot \vec p -
\overline A M A} 
\quad .
\label{genprop}
\eeq
These rules can be used to generate the relevant S-matrix elements 
for any given cross section yielding the transition probability
per unit volume per unit time.
This probability is dependent upon the normalization of the 
incident beams and must be divided by a factor $F$ to account for the
properties of the initial state and yield a physical cross section.

Consideration of two colliding beams (not necessarily collinear)
motivates the definition
\beq
F = N_1 N_2 |\vec v_1-\vec v_2| 
\quad ,
\label{F}
\eeq
in terms of the beam densities, $N_1$ and $N_2$, and the magnitude of
the beam velocity difference.

In the absence of Lorentz violation, this factor may be written in the
Lorentz-covariant form 
\beq
F=4[(p_1 \cdot p_2)^2 - m^4]^{1 \over 2}
\quad ,
\eeq
valid in any reference frame.
In the Lorentz-violating case, a complication is encountered since the
field redefinition $\psi = A \chi$ is {\it frame dependent}.
This means that the physical states will appear to different observers
with fundamentally different properties.  
For example, the observed mass of an electron with a $c$ coupling term
(shown in the next section) is
$\tilde m
\equiv m(1 - c_{00})$ which depends on observer reference frame through
the frame-dependent quantity $c_{00}$.

To circumvent difficulties associated with the field redefinition we
have found it most convenient to do the entire calculation of any given
cross section in a single observer reference frame.
This involves calculation of the flux factor in the same frame that the
S-matrix element was computed in.
The velocities of the beams are calculated using the general 
expression for the group velocity of a wave packet 
\beq
\vec v_g \equiv \vec \nabla_{p} E(\vec p) 
\quad .
\label{vg}
\eeq
Note that the modified dispersion relation may in general cause $E$ to
depend on the direction of $\vec p$ yielding a velocity that is not
parallel to the momentum. 

\section{Relativistic $e^- e^+$ Physics}

To gain more insight into the specifics of the general procedure 
given in the previous sections, attention will be restricted to
ultrarelativistic electron and positron physics. The relevant energy scale
here is taken to be that of a  high-energy collider, still much lower than
the scale where  causality or stability of the low-energy effective
theory come into question.~\cite{kle}

The full QED lagrangian in Eq.(\ref{baselag}) simplifies in the
relativistic limit.
The derivative couplings $c$ and $d$ will dominate over the 
nonderivative $a$, $b$, and $H$ couplings at high energies and
momenta.  
Moreover, if the beams are unpolarized, the effects of the $d$
terms will average out in the sum over right- and left-handed 
particles.
In short, only $c$ will contribute to ultrarelativistic, unpolarized
scattering experiments.

The lagrangian of Eq.(\ref{baselag}) therefore reduces to
\beq
\cl = \half i (\et_{\mu\nu} +  c_{\mu\nu}) \overline{\ps}
\ga^{\mu}  \lrDnuupper \ps
- m \overline{\ps} \ps 
\quad ,
\label{pslag}
\eeq
in the above limit.
The field redefinition in Eq.(\ref{fredef}) used
to eliminate  the time derivatives takes the specific form
(to lowest order in $c$)
\beq
\psi \equiv A \chi = (1 - \half  c_{\mu 0} \gamma^{0}
\gamma^{\mu})\chi
\quad .
\eeq
The lagrangian  expressed in terms of the redefined field $\chi$ 
becomes
\beq
\cl = 
\half i (\et_{\mn} +  \cC_{\mn}  )  \overline \chi
\gamma^{\mu}\lrDnuupper \chi  - \tilde{m} \overline \chi \chi
\eeq
with the definitions
\bea
\tilde m &\equiv& m(1 -  c_{00} )
\quad ,
\nonumber\\
\cC_{\mn}  &\equiv &
 c_{\mu\nu}  -  c_{\mu 0}  \et_{0 \nu} 
+  c_{\nu 0}  \et_{0 \mu} -  c_{00}  \et_{\mu\nu}
\quad ,
\eea
or, in matrix form,
\beq
 \cC  = \left(  
\begin{array}{cccc}
 0 &  c_{01} + c_{10} & c_{20} + c_{02} & c_{30} + c_{03} \\
 0 &  c_{11} + c_{00} & c_{12} & c_{13} \\
 0 &  c_{21} & c_{22} + c_{00} & c_{23} \\
 0 &  c_{31} & c_{32} & c_{33} + c_{00} \\
\end{array}
 \right)
\quad .
\eeq
Note that the first column of the matrix is zero showing explicitly the
removal of the time derivative couplings.

The next step is to quantize the field $\chi$ and define the 
single-particle states for the theory.
The approach here follows a construction previously presented in 
the literature.~\cite{kle}
First, the relativistic quantum mechanics is constructed
by solving for the dispersion relation and the spinors.
Following this, quantization conditions are imposed on the 
solutions to yeild a positive definite hamiltonian.

The modified Dirac equation for $\chi$ can be solved exactly using
the plane-wave solutions for particles and antiparticles of
\beq
\chi(x) = e^{- i p_{\mu} x^{\mu}} u({\vec p}) \quad ; \quad 
\chi(x) = e^{i p_{\mu} x^{\mu}} v(\vec{p})
\quad .
\label{soln}
\eeq
Focusing on the particle spinor $u(\vec p)$, 
it is found to satisfy
\beq
[(\eta_\mn -  \cC_\mn  ) \gamma^{\mu} p^{\nu} - \tilde m ]
u(\vec{p})  = 0 
\quad .
\eeq
A nontrivial solution implies the dispersion relation
\beq
(p^\mu +  \cC^\mu_{~\nu}  p^\nu) (p_\mu +  \cC_{\mu
\al}
 p^\al)  -
\tilde{m}^{2} = 0 
\quad ,
\eeq
which yields (to lowest order in  $\cC$)
\beq
E(\vec p) \approx 
\sqrt{\vec p^2 + \tilde m^2}
-\fr {p^j  \cC_{jk}  p^k}
{\sqrt{\vec p^2 + \tilde m^2}}
-  \cC^0_{\pt{0}j}  p^j \quad .
\label{emom}
\eeq
The same development applies to the spinors $v(\vec p)$ and yields 
the same dispersion relation.
Therefore the energy is degenerate for particles and antiparticles.
It can also be seen from Eq.(\ref{emom}) that the energy depends 
explicitly on the direction of $\vec p$ allowing the group velocity 
to have a different direction than the momentum.

The free-field theory is constructed by expanding the field 
$\chi(x)$ in terms of fourier components and promoting the 
amplitudes to operators as in the conventional case:
\beq
\chi(x) = 
\int \fr {d^3 \vec{p}} {(2 \pi)^{3} N(\vec p)}
\sum_{\al = 1}^{2} \left[
b_{(\al)}(\vec p) e^{-i p \cdot x} u^{(\al)}(\vec p) 
+ d_{(\al)}^{\dagger}(\vec p) e^{i p \cdot x} v^{(\al)}(\vec p) 
\right] ,
\eeq
where the spinors are normalized to
\bea 
u^{(\al) \dagger} (\vec{p}) u^{(\al^{\prime})}(\vec{p}) =
\de^{\al \al^{\prime}} N(\vec p) 
& , & \quad
v^{(\al) \dagger} (\vec{p}) v^{(\al^{\prime})}(\vec{p}) =
\de^{\al \al^{\prime}} N(\vec p) ,
\nonumber \\
u^{(\al) \dagger} (\vec{p}) v^{(\al^{\prime})}(-\vec{p}) = 0
& , & \quad
v^{(\al) \dagger} (-\vec{p}) u^{(\al^{\prime})}(\vec{p}) = 0 .
\label{orthonorm}
\eea
Quantization is implemented by imposing 
\bea
\{b_{(\al)} (\vec{p}), b^{\dagger}_{(\al^{\prime})}
(\vec{p}^{~\prime}) \} & = & (2 \pi)^3 
N(\vec p) 
\de_{\al \al^{\prime}}
\de^3 (\vec{p} - \vec{p}^{~\prime}) ,
\nonumber \\
\{d_{(\al)} (\vec{p}), d^{\dagger}_{(\al^{\prime})}
(\vec{p}^{~\prime}) \} & = & (2 \pi)^3 
N(\vec p)
\de_{\al \al^{\prime}}
\de^3 (\vec{p} - \vec{p}^{~\prime}) ,
\eea
on the mode operators.
Translational invariance implies that there is a conserved 
energy-momentum tensor explicitly given by
\beq
\Theta^{\mn} = 
\fr i 2 
\tilde{\et}_{\al}^{\ \mu} 
\overline \chi 
\gamma^{\al} \lrprtnu \chi .
\eeq
The corresponding conserved four-momentum is diagonal
in the creation and annihilation operators:
\bea
P^{\mu} & = & \int d^{3} \vec x : \Theta^{0\mu} : 
\nonumber \\
& = & \int \fr {d^{3} \vec p} {(2 \pi)^{3} N(\vec p)} 
p^{\mu}\sum_{\al = 1}^{2} \left[ b_{(\al)}^{\dagger}(\vec p) 
b_{(\al)}(\vec p) + d_{(\al)}^{\dagger}(\vec p) d_{(\al)}(\vec p) 
\right] .
\eea
Note that this would not have been the case if the field redefinition
had not been implemented.

The single particle states can be defined in the conventional manner using 
the mode operators acting on the vacuum state.  
The resulting normalization for these states is $\langle
p^\prime,\al^\prime | p,
\al
\rangle  = (2 \pi)^3 N(\vec p)
\de_{\al \al^\prime}
\de^3(\vec p^\prime - \vec p)$.
It follows that the number density for an incident plane wave is $N(\vec
p)$ particles per unit volume.

\section{Cross Section for $e^- e^+ \rightarrow 2 \gamma$}

In this section the theory that has been developed for relativistic
electron-positron physics is applied to the explicit process of pair
annihilation into two photons. 
The explicit cross section is obtained and various properties of the 
Lorentz-violating effects are discussed.

The relevant modifications to the Feynman rules for this case are the 
modified fermion propagator
\beq
S_F(p) ={ i \over p_\mu (\ga^\mu +  C_\nu^{~\mu}  \ga^\nu - \tilde m)}
\quad ,
\eeq
and the modified vertex factor of
\beq
-i e (2 \pi)^4 \delta^4(\sum p) [\ga^\mu
+ \cC_\nu^{~ \mu} \gamma^\nu]
\quad ,
\eeq
arising from the modified lagrangian.

The tree-level diagrams for the process $e^+ e^- \rightarrow 2\ga$ are
the same as in the conventional case, with the modified propagator and
vertex factors included.
The resulting S-matrix element is
\bea
S_{fi} &=& -ie^{2}(2\pi)^{4} \de^4 (k_1 + k_2 - p_1 - p_2) 
\overline{v}(p_{2}) \biggl[ {\slash{\tilde\ep_{2}}} \fr 1 
{{\slash{\tilde p_{1}}} - 
{\slash{\tilde k_{1}}} - \tilde m} {\slash{\tilde\ep_{1}}} + (1
\leftrightarrow 2)\biggr] u(p_{1}) .
\nonumber \\
\eea
In this equation,
$p_1$, $p_2$ are the electron and positron momenta,
while $k_1$, $k_2$ are the photon momenta.
The spinors $u$, $v$ solve the modified Dirac equation
after the reinterpretation,
while $\ep_1$, $\ep_2$ are the two photon polarization vectors.
The notation ${\slash{\tilde p}} \equiv (\eta_{\mn} + \cC_{\mn})\ga^\mu
p^\nu$ is used to simplify the expression. In working with
the above expression, it is important to realize that the electron and
positron energies satisfy  modified dispersion relations and the spinors
are exact solutions to the modified Dirac equation satisfied by $\chi$.

To define the physical cross section, the factor $F$ in Eq.(\ref{F})
must be calculated in the same frame that the above S-matrix element is 
evaluated in.
We choose the center of momentum frame for the evaluation of both of these
quantities.  
Note that the group velocities of the beams are not
necessarily equal and opposite in this frame due to the modified
dispersion relations 
as is illustrated in figure 1.
The scattering angle is defined using the incoming electron momentum and 
the outgoing photon momentum as $\cos{\theta} = \hat p \cdot \hat k$.
The magnitude of the velocity difference in the center of momentum
frame in the relativistic limit  is 
$|\vec v_1 - \vec v_2| \approx 2(1 - \hat p_j
\cC^{jk}\hat p_k)$.

\begin{figure} 
\centerline{\psfig{figure=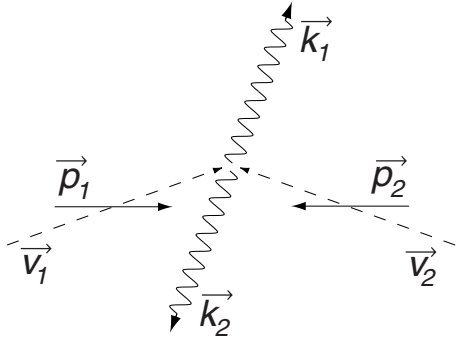,width=0.5\hsize}}
\smallskip
\caption{
Kinematics for the process $e^- e^+ \rightarrow 2 \ga$ 
in the center of momentum frame.
Note that the velocities of the incident
beams are not in general equal and opposite due to the difference in
directions of the group velocity $\vec v$ and the momentum $\vec p$. }
\label{fig1}
\end{figure}

To obtain the physical cross section, the conventional steps are 
now followed.
The electron and positron spins are averaged over and the final state 
photon polarizations are summed over.
The momentum of the electron beam is chosen to point in the 3-direction
for simplicity.
The incident flux factor is divided out and the final state photon phase
space is included.
As a final step, the azimuthal angle is integrated over to simplify the 
expression.

The resulting cross section becomes
\beq
{d \si \over d \cos{\th}} = \int_0^{2 \pi} d \phi {d \si \over d \Om} = 
\left( {d \si \over d \cos{\th}} \right)_{QED}
[1 + \De]
\quad ,
\eeq 
where the first factor represents the conventional QED cross section 
and the Lorentz-violating correction $\De$ is given by
\beq
\De = c_{00} + c_{33} - 2 {\cos^2{\theta} \over 1 + \cos^2{\th}}
(c_{11} + c_{22} - 2c_{33})
\quad .
\eeq
The correction consists of a piece that is an overall scaling of the
cross section and another part that modifies the angular dependence.
Note that the cross section is now explicitly time-dependent since the 
$c_{ij}$ components change as the Earth rotates.
This occurs because the 3-direction points along the electron beam
momentum that rotates along with the earth.

To understand the time dependence, it is useful to transform to a fixed 
basis $(\hat X, \hat Y, \hat Z)$ with the $\hat Z$-direction along the
axis of the earth and the other two directions fixed with respect to
the background stars.~\cite{kl}
The components of $c$ with respect to this basis are fixed so the 
time-dependence can be explicitly extracted from the cross section.
For example, the term $c_{33}$ expressed in terms of the fixed basis is
\bea
c_{33} = 
&& 
c_{ZZ} 
+ \half (c_{XX} + c_{YY} - 2 c_{ZZ}) \sin^2{\chi} 
\nonumber \\
&&
+ \half (c_{YZ} + c_{ZY})\sin{2\ch} \sin{\Omega t} 
+ \half (c_{XZ} + c_{ZX})\sin{2\ch} \cos{\Omega t}
\nonumber \\
&& 
+ \half (c_{XY} + c_{YX})\sin^2{\ch} \sin{2\Omega t}
+ \half (c_{XX} - c_{YY})\sin^2{\ch} \cos{2\Omega t} .
\eea
In this expression, $\chi$ is the angle between $\hat p$ and $\hat Z$,
and $\Om$ is the sidereal rotation frequency of the earth.
Note that there are three components to the time dependence:
a time-independent factor, a piece that varies with period $T = 2 \pi /
\Om$ and another piece with period $T = \pi / \Om$.
The conventionally measured cross section gives the time integrated
constant part as the times are effectively averaged.

\section{Summary}

A framework has been developed for calculation of cross sections 
and decay rates within the context of the Lorentz-violating
standard model extension.
The modified Feynman rules and kinematical factors have been  deduced.
It was found that it is most convenient to perform the analysis 
in a single frame due to complications arising from the frame-dependent
field redefinition.
The cross section for $e^- e^+ \rightarrow 2 \ga$ was calculated 
explicitly using the framework developed.
It was found that the modifications to the conventional
cross section are dependent on sidereal time due to the rotation of the
earth. This time dependence is a qualitatively new feature of cross
sections in the presence of Lorentz violation.

\section*{Acknowledgments}
This work was supported in part by a University of South Florida 
Division of Sponsored Research grant.

\end{document}